\documentclass[12pt,reqno]{amsart}

\usepackage{amsmath, amssymb, amsthm, graphicx, hyperref}

\newtheorem{theorem}{Theorem}[section]
\newtheorem{conjecture}[theorem]{Conjecture}

\theoremstyle{remark} 

\title{\boldmath Relating Hodge Atoms, Spectral Triples,\\and BPS
  Flows}

\author{Mark Raugas} \address{Pacific Northwest National Laboratory,
  1100 Dexter Ave N, Seattle, WA 98109} \email{raugas@pnnl.gov}

\subjclass[2020]{Primary 81T30; Secondary 14E08, 14D05, 14J35
  } \keywords{Hodge atom, cyclic cohomology, spectral
  triple, wall crossing, Kuznetsov component, BPS solitons, birational
  invariance, Landau-Ginzburg mirror}
\begin{document}

\begin{abstract}
  We compare algebraic and analytic pictures relevant to the study of
  birational invariants.  Motivated by recent advances in the
  development of non-commutative Hodge structures, we examine their
  implication for quiver gauge field theory on the cubic fourfold.  By
  interpreting the semiorthogonal property as a dynamical selection
  rule, we conjecture that the K3 Hodge atom of the cubic fourfold
  represents a protected quantum phase whose spectra remain invariant
  under non-perturbative tunneling processes.
\end{abstract}

\maketitle \flushbottom

\section{Introduction}

The classification of algebraic varieties up to birational equivalence
is one of the central enterprises of modern algebraic geometry. While
classical invariants such as the fundamental group, plurigenera, and
classical Hodge numbers are highly effective in classifying curves and
surfaces, they frequently fail to distinguish birationally
inequivalent varieties in higher dimensions. This limitation is
famously embodied by the rationality problem of the cubic fourfold:
while the rationality question is settled for cubic hypersurfaces of
dimension at most three \cite{Clemens1972}, the case of the cubic
fourfold until recently has resisted resolution.

Modern approaches replace the study of classical geometric spaces with
the study of their bounded derived categories of coherent sheaves and
their deformations via Gromov-Witten theory. In the framework of
non-commutative (nc) Hodge structures of Katzarkov,
Kontsevich, and Pantev \cite{KKP2008}, the quantum
differential equation (the Dubrovin connection) over the moduli space
of quantum parameters is used to extract rigid categorical data called Hodge
atoms from the Stokes matrices at irregular singularities
\cite{KKPY2025}. Because these atoms behave additively under the
semiorthogonal decompositions induced by blow-ups, they provide
birational invariants capable of probing varieties whose moduli
exhibit irregular singularities.

We review some of the machinery involved and draw parallel to other formalisms.  We begin with a quick review of relevant material from
Algebraic Geometry, including F-bundles, quantum
      multiplication, and the Kuznetsov components of derived
      categories to provide the categorical framework in which Hodge
      atoms are defined and their birational invariance is
      established.  Then we examine the same spaces using
      Connes' cyclic
      cohomology \cite{Connes1994}, realized through Witten-deformed
      Dolbeault spectral triples on Landau-Ginzburg mirrors.
      This is with the aim towards supplying a potential path towards
      an analytic construction of similar invariants via
      heat kernel localization.
      Seiberg-Witten $u$-plane
      duality \cite{Seiberg1994}, BPS wall-crossing \cite{GMN2013},
      and split attractor flows \cite{Denef2000} are reviewed
      to provide geometric
      intuition for the Stokes multipliers and a possible dynamical
      interpretation of categorical rigidity in supersymmetric
      field theories.
      By developing these three pictures in parallel, we aim to relate how
      birational surgery manifests in each formalism.

      We then examine physical interpretations in the case of the cubic fourfold
from semi-classical approximation
to the role of instanton-mediated tunneling between vacuum
states.
By framing the semiorthogonal property of the Kuznetsov
component as a dynamical selection rule, we argue that the K3 Hodge
atom of the cubic fourfold
is protected from tunneling processes associated with birational
surgery due to full acyclicity of relevant RHom complexes.

\section{Background}

\subsection{From Static Hodge Theory to F-bundles}
In classical geometry, the Hodge filtration $F^\bullet$ on $H^*(X,
\mathbb{C})$ relates Betti cohomology to De Rham cohomology. In the
non-commutative setting of differential graded (DG) categories
\cite{Keller2006}, these static decompositions are replaced
with vector bundles over a formal disk.

The ``F-bundle'' is a Variation of Hodge Structure (VHS) associated
with quantum cohomology. For a smooth projective variety $X$, consider
the trivial bundle $\mathcal{H} = H^*(X, \mathbb{C}) \times
\mathbb{C}_z$ over the complex plane $\mathbb{C}_z$, where $z$ is the
gravitational descendant parameter. The bundle is equipped with the
Dubrovin (or Quantum) connection \cite{Dubrovin1996}:
\begin{equation}
    \nabla_z = \frac{d}{dz} - \frac{1}{z} (c_1(X) \star) -
    \frac{1}{z^2} (\mu)
\end{equation}
where $\star$ denotes quantum multiplication and $\mu$ is the grading
operator. The F-bundle is the pair $(\mathcal{H}, \nabla_z)$. Unlike
Connes' spectral triples which probe metric geometry via a Dirac
operator $D$, non-commutative Hodge theory uses this connection on
periodic cyclic homology \cite{Loday1992, Iritani2009} to probe complex/algebraic
geometry. The birational invariants (called Hodge Atoms) then arise from the rigid
spectral components of the irregular singularity at $z=0$.

\subsection{U-plane Duality and Mirror Symmetry}
The Variation of Hodge Structure (VHS) argument for F-bundles and Witten's
$u$-plane duality are manifestations of the geometry of flat
connections over a moduli space, governed by Homological Mirror
Symmetry \cite{Kontsevich1994, Auroux2007}.

In Seiberg-Witten theory, the vacuum moduli space is parameterized by
$u = \langle \operatorname{Tr}(\phi^2) \rangle$. The physics is
determined by periods of a meromorphic differential on a family of
elliptic curves, providing a Weight 1 VHS governed by the classical
Gauss-Manin connection. Through geometric engineering, compactifying
Type IIA string theory on a non-compact Calabi-Yau threefold mirrors
this setup: the Dubrovin connection of the F-bundle on the A-side is
dual to the Gauss-Manin connection on the Seiberg-Witten curve.

Singularities play identical roles: massless monopoles at $u = \pm
\Lambda^2$ in Seiberg-Witten theory are the irreducible rigid data of
the vacuum, just as Hodge atoms are the rigid spectral data of the
non-commutative space.
The Kontsevich-Soibelman wall-crossing formula \cite{Kontsevich2008}
bridges the sudden jumps in the BPS spectrum across curves of marginal
stability with the Stokes phenomenon of asymptotic expansions in the
F-bundle.

\subsection{The Calabi-Yau Quintic and Wall-Crossing}
By examining the
case of the quintic threefold $X \subset \mathbb{P}^4$ one
is able to understand
how Kontsevich-Soibelman (KS) wall-crossing formula translates BPS
states into Stokes data.
BPS states, modeled as objects in $D^b(X)$ with charge $\gamma$,
possess a central charge $Z_\gamma(z) = \int_\gamma \Omega_z$. A wall
of marginal stability occurs where $\arg(Z_{\gamma_1}) =
\arg(Z_{\gamma_2})$. Crossing this wall causes the Donaldson-Thomas
invariant $\Omega(\gamma)$ to jump. The KS formula dictates that the
radially ordered product of formal symplectomorphisms remains
invariant:
\begin{equation}
    \prod_{\text{phase order}} U_\gamma^{\Omega(\gamma)_{z_+}} =
    \prod_{\text{phase order}} U_\gamma^{\Omega(\gamma)_{z_-}}
\end{equation}
The Stokes rays of the exact WKB analysis of the quantum connection
are identical to these walls of marginal stability, and the Stokes
matrices patching asymptotic solutions are precisely the KS operators.

While Calabi-Yau manifolds ($c_1=0$) yield regular singularities,
(such as in the complex moduli space of the mirror
quintic $X^\vee$, the Large Complex
Structure Limit ($z=0$) and the conifold point ($z=5^{-5}$)),
Fano
varieties such as the cubic fourfold (with $c_1 > 0$) introduce an irregular singularity at $z=0$ in
the Dubrovin connection due to presence of a $\frac{1}{z}(c_1 \star)$ factor.

Katzarkov, Kontsevich, Pantev, and Yu proved that Stokes
data at this irregular singularity is additively compatible with
blow-ups. Under a blow-up, Stokes data splits into ``flexible'' atoms
(generated by the blow-up) and ``rigid'' atoms and factoring out the
flexible components yields a birational invariant. This provides a
method to test the rationality of the cubic fourfold: if rational, a
cubic fourfold would be birational to $\mathbb{P}^4$, which possesses
only trivial Hodge atoms. By demonstrating that the cubic fourfold
possesses a non-trivial, rigid K3-like Hodge atom, they establish that the
general cubic fourfold is not rational.

\section{The Kuznetsov Component and Categorical Rigidity}
To analyze birational invariants for Fano varieties,
the Kuznetsov component and its associated Serre functor are used.

\subsection{Categorical Filtering and Semiorthogonal Decompositions}
For a smooth projective variety $X$, the bounded derived category
$D^b(X)$ often admits a semiorthogonal decomposition. In the case of
the cubic fourfold $X \subset \mathbb{P}^5$, one can utilize an
exceptional collection of line bundles to define the Kuznetsov
component $\mathcal{A}_X$:
\begin{equation}
    D^b(X) = \langle \mathcal{A}_X, \mathcal{O}_X, \mathcal{O}_X(1),
    \mathcal{O}_X(2) \rangle
\end{equation}
where $\mathcal{A}_X$ is the right orthogonal complement:
\begin{equation}
    \mathcal{A}_X = \{ A \in D^b(X) \mid
    \operatorname{RHom}(\mathcal{O}_X(i), A) = 0, \, \forall i \in
    \{0,1,2\} \}
\end{equation}

\noindent For the cubic fourfold, $\mathcal{A}_X$ is a
geometric K3 category, satisfying:
\begin{equation}
    \mathbf{S}_{\mathcal{A}_X} \cong [2]
\end{equation}
\noindent where $\mathbf{S}_{\mathcal{A}_X}$ is the Serre functor.  This identifies $\mathcal{A}_X$ as a rank-24 irregular
$\mathcal{D}$-module in the F-bundle. Unlike the isolated critical
points associated with the line bundles, the K3 atom corresponds to a
rigid, degenerate critical locus in the Landau-Ginzburg mirror
\cite{Addington2014, Kuznetsov2007}.

\subsection{HKR and the JLO Cocycle}
The Hochschild-Kostant-Rosenberg (HKR) isomorphism relates this
categorical structure to the analytic index via Hochschild homology
$HH_\bullet(\mathcal{A}_X)$. The JLO cyclic cocycle $\tau_W$ provides
the analytic representative of the Chern character in the
Witten-deformed spectral triple:
\begin{equation}
    \tau_W(a_0, \dots, a_n) = \int_{\Delta_n} \operatorname{Tr}_s
    \left( a_0 e^{-t_0 D_W^2} [D_W, a_1] \dots [D_W, a_n] e^{-t_n
      D_W^2} \right) dt
\end{equation}
The Serre functor shift $\mathbf{S}_{\mathcal{A}_X} \cong [2]$ ensures that the
JLO cocycle localizes onto the K3 critical locus, preserving the Mukai
pairing through birational surgery \cite{GetzlerSzenes1989, Mukai1987}.

\subsection{Quantum Blow-up Theorem and Weight Filtration}
The birational invariance of the K3 atom is formalized by the Quantum
Blow-up Theorem. In the monodromy weight filtration $W_\bullet$ of the
F-bundle, the blow-up of a center $Y$ induces an additive splitting:
\begin{equation}
    W_\bullet(\mathcal{H}_{\tilde{X}}) \cong
    W_\bullet(\mathcal{H}_{X}) \oplus \bigoplus_{j}
    W_\bullet(\mathcal{H}_{Y} \otimes \mathbb{L}^j)
\end{equation}
This splitting ensures that the rigid $W_\bullet$ associated with the
K3 atom is protected from the components arising from birational surgery,
providing an obstruction to rationality for the
cubic fourfold.
\subsection{Numerical Invariants and the Mukai Lattice}
The definitive obstruction to rationality for the cubic fourfold is
contained in the numerical invariants of the Mukai lattice
$\tilde{H}(\mathcal{A}_X, \mathbb{Z})$. This lattice, derived from the
numerical Grothendieck group $K_0(\mathcal{A}_X)$, is equipped with
the Mukai pairing:
\begin{equation} 
  \langle v(E), v(F) \rangle = - \chi (E,F) = -\sum (-1)^i \dim \operatorname{Ext}^i(E, F)
\end{equation}
In the Landau-Ginzburg mirror, this pairing corresponds to
the intersection pairing of Lefschetz thimbles, which determine the
entries of the K3-type Stokes block $S_{\mathcal{A}}$.

While $\mathcal{A}_X$ and a geometric K3 surface $S$ share the same
Serre functor $\mathbf{S} \cong [2]$, their Mukai lattices are generally not
isomorphic. The K3 atom $\mathcal{A}_X$ possesses a lattice signature
that is protected by the semiorthogonality of the Kuznetsov
decomposition. Under birational surgery, the Quantum Blow-up Theorem
ensures that new exceptional objects $E_j$ possess trivial Mukai
vectors that are orthogonal to
$\mathcal{A}_X$.

\subsection{Spectral Networks and the Geometry of Vacua}
To derive the numerical invariants of the K3 atom, Gaiotto-Moore-Neitzke (GMN) spectral networks \cite{GMN2013, GMN2013b, GMN2009} can be used
to map the intersection
pairing of Lefschetz thimbles. In the Landau-Ginzburg mirror, these
networks are formed by the union of BPS rays emanating from the
critical points $p_i$ of the superpotential $W$. The off-diagonal
Stokes multipliers $S_{ij}$ are then recovered by counting the oriented
gradient flow lines connecting vacua in the $W$-plane:
\begin{equation}
    S_{ij} = \langle \Delta_i, \Delta_j \rangle =
    \chi(\operatorname{RHom}(E_i, E_j))
\end{equation}

The K3 atom is then represented by a dense sub-network of thimble
intersections within the spectral network
that recovers the Mukai pairing $( \cdot, \cdot )_K$ and
the rank-24 lattice signature.  The semiorthogonal ordering of the Kuznetsov
decomposition ensures that no BPS rays connect vacua associated to birational
surgery back to the K3 cluster.

Blowing up along a center $Y$ expands the derived category via Orlov's
semiorthogonal decomposition \cite{Orlov1992}, $D^b(\tilde{X}) =
\langle D^b(X), D^b(Y)_1, \dots \rangle$, affecting the stability
manifold and Donaldson-Thomas (DT) invariants.
In the B-Model, we can project to a
      rigid subcategory of Kuznetsov Components.  For the cubic fourfold, this
      is the K3-like component $\mathcal{A}_X$ in the decomposition
      $D^b(X) = \langle \mathcal{A}_X, \mathcal{O}, \mathcal{O}(1),
      \mathcal{O}(2) \rangle$.  In the A-Model, the quantum
      D-module of a blow-up splits as a direct sum of the original
      D-module and shifted copies of the center's D-module.
Hodge Atoms are then the spectra of these irreducible D-modules.

\subsection{BPS Wall-Crossing and Global Stability}
The Stokes data derived from the irregular singularity at $z=0$ is not
static; it varies across the moduli space of the variety according to
the Kontsevich-Soibelman (KS) wall-crossing formula. In the
Landau-Ginzburg mirror, walls of marginal stability occur where the
phases of the central charges $Z_\gamma = \int_\gamma e^{W/z} \Omega$
for different BPS states align.

Crossing these walls induces a jump in the Stokes multipliers $S_{ij}$
through the Stokes phenomenon. However, the KS formula ensures that
the total non-commutative Hodge structure remains invariant:
\begin{equation}
    \prod_{\gamma \in \text{rays}} U_\gamma^{\Omega(\gamma)_{+}} =
    \prod_{\gamma \in \text{rays}} U_\gamma^{\Omega(\gamma)_{-}}
\end{equation}
where $U_\gamma$ are the automorphisms of the quantum torus.

For the cubic fourfold, this wall-crossing behavior demonstrates the
rigidity of the K3 atom. While the flexible thimbles associated with
the line bundles may undergo mutations, the semiorthogonal ordering of
the Kuznetsov component $\mathcal{A}_X$ prevents these transitions
from destabilizing the K3 sub-network.

\section{Explicit Spectral Triples for LG Mirrors}
Landau-Ginzburg (LG) models are
equipped with a complex structure from the B-model
and from the Dolbeault complex one can construct associated Witten-deformed
Dirac operators.\cite{Witten1982}

For an LG model on a K\"{a}hler manifold $X$ with holomorphic
superpotential $W: X \to \mathbb{C}$, the non-commutative space is
captured by a spectral triple $(\mathcal{A}, \mathcal{H}, D_W)$. We
set the algebra to $\mathcal{A} = C_c^\infty(X)$ and the Hilbert space
to the $L^2$-completion of the anti-holomorphic differential forms,
$\mathcal{H} = L^2(X, \Lambda^{0,\bullet} T^* X)$. The Witten-deformed
Dolbeault Dirac operator is then given by:
\begin{equation}
    D_W = \bar{\partial} + \bar{\partial}^* + s(\partial W \wedge
    \cdot) + s(\iota_{\overline{\partial W}})
\end{equation}
where $s = 1/z$. The square of this operator introduces a scalar
potential governing the spectrum:
\begin{equation}
    D_W^2 = \Delta_{\bar{\partial}} + s^2 |\partial W|^2 +
    s(\text{Hessian terms})
\end{equation}
As $z \to 0$ (thus $s \to \infty$), the heat kernel $\exp(-t D_W^2)$
localizes around the potential wells where $\partial W = 0$.

\subsection{The Spectral Triple for $\mathbb{P}^1$}
The mirror to the complex projective line $\mathbb{P}^1$ is the
punctured plane $X = \mathbb{C}^\times$ with coordinate $x$ and
superpotential $W(x) = x + q/x$.
\begin{itemize}
    \item \textbf{Algebra:} $\mathcal{A} =
      C_c^\infty(\mathbb{C}^\times)$.
    \item \textbf{Hilbert Space:} $\mathcal{H} =
      L^2(\mathbb{C}^\times, \Lambda^{0,\bullet} T^*
      \mathbb{C}^\times)$.
    \item \textbf{Localization Potential:} The dominant term governing
      the heat kernel asymptotics is $s^2 |\partial W|^2 = s^2 \left|
      1 - \frac{q}{x^2} \right|^2$.
\end{itemize}
The potential strictly vanishes at $x = \pm \sqrt{q}$. In Connes'
theory, the heat kernel's asymptotic expansion around these two points
yields the JLO cyclic cocycle \cite{JLO1988}. Geometrically, these two
discrete localizations correspond to the two Hodge Atoms of
$\mathbb{P}^1$, mirroring the exceptional objects $\mathcal{O}$ and
$\mathcal{O}(1)$ generating $D^b(\mathbb{P}^1)$.

\subsection{The Spectral Triple for $\mathbb{P}^2$}
The mirror to the projective plane $\mathbb{P}^2$ scales this
construction to two dimensions. The space is the algebraic torus $X =
(\mathbb{C}^\times)^2$ with coordinates $(x,y)$ and superpotential
$W(x,y) = x + y + q/(xy)$.
\begin{itemize}
    \item \textbf{Algebra:} $\mathcal{A} =
      C_c^\infty((\mathbb{C}^\times)^2)$.
    \item \textbf{Hilbert Space:} $\mathcal{H} =
      L^2((\mathbb{C}^\times)^2, \Lambda^{0,\bullet} T^*
      (\mathbb{C}^\times)^2)$.
    \item \textbf{Localization Potential:} The dominant term governing
      the heat kernel asymptotics is $s^2 |\partial W|^2 = s^2 \left(
      \left| 1 - \frac{q}{x^2y} \right|^2 + \left| 1 - \frac{q}{xy^2}
      \right|^2 \right)$.
\end{itemize}
This potential vanishes precisely at the three points $x = y = q^{1/3}
e^{2\pi i m / 3}$ for $m \in \{0, 1, 2\}$ and the heat kernel splits into
three local contributions. The Ext-cohomology between these three
atoms models the gradient flows (instantons) tunneling
between these three potential wells.

\subsection{Del Pezzo Surfaces: Blowing up $\mathbb{P}^2$}
To analytically observe birational surgery, consider the first few del Pezzo
surfaces formed by blowing up $\mathbb{P}^2$ at $k$ points in general
position ($k=1, 2, 3$). These spaces are Fano, and while their complex
structure is rigid, their K\"ahler moduli space exhibits an irregular
singularity in the large volume limit.

Algebraically, blowing up $k$ points adds $k$ exceptional objects to
the derived category via Orlov's formula:
\begin{equation}
    D^b(\operatorname{Bl}_k \mathbb{P}^2) = \langle \mathcal{O},
    \mathcal{O}(1), \mathcal{O}(2), E_1, \dots, E_k \rangle
\end{equation}

Analytically, the B-model reflects this addition. Under the
Witten-deformed Dolbeault operator $D_{W_k}$, the mirror space $X_k$
yields a localization potential $s^2 |\partial W_k|^2$. While the
original $\mathbb{P}^2$ potential possessed exactly three critical points,
the deformed superpotential $W_k$ has $3 + k$
non-degenerate critical points and as $s \to \infty$, the heat kernel $\exp(-t D_{W_k}^2)$ localizes into
$3+k$ discrete potential wells. The JLO cyclic cocycle cleanly splits
into contributions from the 3 original vacua (which remain unperturbed
locally) plus $k$ new, distinct localized atoms. This is an indication
that Hodge atoms behave strictly additively under birational morphisms.

\section{Stokes Multipliers and Split Attractor Flows}
In the LG mirror $(X, W)$, critical
points $p_i$ correspond to vacua \cite{Denef2000, Denef2008}.  Each vacuum corresponds to an
irreducible Hodge atom or an exceptional object $E_i$ in the derived
category $D^b(X)$.  The Stokes multiplier $S_{ij}$ represents the
entries of the matrix that describes how the basis of solutions to the
quantum differential equation jumps as one moves in the parameter
space.  For a phase $\theta$, the Lefschetz thimble $\Delta_i(\theta)$
is the stable manifold of the gradient flow lines emanating from the
critical points \cite{Seidel2008}.

A split attractor flow (a BPS soliton) exists when a gradient flow
line connects $p_i$ and $p_j$ at the critical angle $\theta_{ij} =
\arg(\lambda_i - \lambda_j)$, where $\lambda$ is a critical value of
the superpotential.  Crossing this angle induces a Picard-Lefschetz
jump:
\begin{equation}
    \Delta_i(\theta_+) = \Delta_i(\theta_-) \pm \langle \Delta_i,
    \Delta_j \rangle \Delta_j(\theta_-)
\end{equation}
The off-diagonal Stokes multiplier $S_{ij}$ is the
intersection number $\langle \Delta_i, \Delta_j \rangle$, equal to
$\chi(\operatorname{RHom}(L_i, L_j))$, where $L_i$ and $L_j$ in the
A-model are Lagrangian branes whose boundary conditions are defined by
the thimbles $\Delta_i$ and $\Delta_j$ and
$\operatorname{RHom}(L_i, L_j)$ describes the morphism arrows of the
Ext-quiver.

\subsection{Split Flows and Multipliers for $\mathbb{P}^1$}
For $\mathbb{P}^1$, the LG mirror has critical values $\lambda_\pm =
\pm 2\sqrt{q}$. The derived category is generated by the exceptional
collection $\langle \mathcal{O}, \mathcal{O}(1) \rangle$. The Stokes
multiplier $S_{12}$ between these two vacua is given by the Euler
characteristic of their Ext complex:
\begin{equation}
    S_{12} = \dim \operatorname{Hom}(\mathcal{O}, \mathcal{O}(1)) =
    \dim H^0(\mathbb{P}^1, \mathcal{O}(1)) = 2
\end{equation}
Physically, this intersection number dictates the existence of
two distinct split attractor flows (BPS solitons) interpolating between
the vacuum at $+\sqrt{q}$ and the vacuum at $-\sqrt{q}$ in the complex
plane.

\subsection{Split Flows and Multipliers for $\mathbb{P}^2$}
For $\mathbb{P}^2$, the superpotential yields three critical values
$\lambda_m = 3 q^{1/3} e^{2\pi i m / 3}$. The derived category is
generated by $\langle \mathcal{O}, \mathcal{O}(1), \mathcal{O}(2)
\rangle$. The Stokes multipliers between adjacent vacua in this
exceptional collection are computed via the global sections:
\begin{equation}
    S_{m, m+1} = \dim \operatorname{Hom}(\mathcal{O}(m),
    \mathcal{O}(m+1)) = \dim H^0(\mathbb{P}^2, \mathcal{O}(1)) = 3
\end{equation}
This establishes that there are three distinct split attractor
flows connecting each pair of adjacent vacua. The resulting spectral
network in the $W$-plane forms an equilateral triangle with 3 arrows
directed along each edge, describing the Markov quiver of
$\mathbb{P}^2$.

\section{Spectral Triples, the Ext-Quiver, and the Cubic Fourfold}
To fully appreciate the irregular nature of the cubic fourfold's Hodge
atoms, it is instructive to contrast its spectral triple with that of
a Calabi-Yau manifold.

\subsection{The Quintic Threefold}
For the quintic threefold $Y \subset \mathbb{P}^4$, the first Chern
class vanishes ($c_1(Y) = 0$). Consequently, the quantum differential
equation possesses only regular singularities. Under mirror symmetry,
the B-model mirror $Y^\vee$ is another Calabi-Yau threefold, and the
Landau-Ginzburg superpotential is identically zero ($W = 0$).

The associated Dolbeault spectral triple $(\mathcal{A}, \mathcal{H},
D)$ relies strictly on the undeformed Dirac operator $D =
\bar{\partial} + \bar{\partial}^*$. Because there is no potential term
$s^2 |\partial W|^2$ to force localization, the heat kernel $\exp(-t
D^2)$ probes the global, continuous geometry of $Y^\vee$. The
resulting cyclic cohomology does not split into discrete atoms,
reflecting the fact that the derived category $D^b(Y)$ is a single,
indecomposable Calabi-Yau category.

\subsection{The Cubic Fourfold and the K3 Atom}
In contrast, the cubic fourfold $X \subset \mathbb{P}^5$ is Fano in
degree $d=3$, and by the adjunction formula, its canonical bundle is
$K_X = \mathcal{O}_X(-3)$. Thus, $c_1(X) = 3H > 0$, implying an {\em
  irregular} singularity. Its bounded derived category admits the
Kuznetsov semiorthogonal decomposition \cite{Kuznetsov2010}:
\begin{equation}
    D^b(X) = \langle \mathcal{A}_X, \mathcal{O}_X, \mathcal{O}_X(1),
    \mathcal{O}_X(2) \rangle
\end{equation}
where the right orthogonal complement $\mathcal{A}_X$ is a K3 category
(its Serre functor acts as $\mathbf{S}_{\mathcal{A}_X} = [2]$).

In the B-model, the LG mirror to the cubic fourfold is equipped with a
non-trivial holomorphic superpotential $W$. The Witten-deformed
Dolbeault operator $D_W$ introduces the localization potential
$s^2|\partial W|^2$. For the cubic fourfold, the critical locus where
$\partial W = 0$ is structured as follows:
\begin{itemize}
    \item \textbf{Isolated Vacua:} There are three isolated,
      non-degenerate critical points. The heat kernel localizes around
      these discrete wells, analogous to the three rank-1 exponential
      $\mathcal{D}$-modules (the trivial atoms) generated by the line
      bundles.
    \item \textbf{The Degenerate Vacuum:} There is a degenerate
      critical locus whose internal geometry mirrors a K3 surface. The
      heat kernel localizes onto this extended locus, yielding a rank
      24 irregular $\mathcal{D}$-module which is the periodic cyclic
      homology $HP_*(\mathcal{A}_X)$ and we associate to the K3 atom
      $\mathcal{A}_X$.
\end{itemize}

The global Stokes matrix $S$ for the quantum differential equation
governs the tunneling between these vacua. It is block
upper-triangular:
\begin{equation}
    S = \begin{pmatrix} S_{\mathcal{A}} &
      \operatorname{Ext}^\bullet(\mathcal{A}_X, \mathcal{O}) &
      \operatorname{Ext}^\bullet(\mathcal{A}_X, \mathcal{O}(1)) &
      \operatorname{Ext}^\bullet(\mathcal{A}_X, \mathcal{O}(2)) \\ 0 &
      1 & 6 & 21 \\ 0 & 0 & 1 & 6 \\ 0 & 0 & 0 & 1
    \end{pmatrix}
\end{equation}
The upper-right off-diagonal blocks encode the Ext-quiver connecting
the K3 atom to the trivial line bundles. Specifically,
$\operatorname{Ext}^1(\mathcal{A}_X,\mathcal{O}_X(i))$ parameterizes
the non-trivial extensions (the physical split attractor flows) that
bind the extended K3 critical locus to the isolated vacua. Because
blowing up $X$ only alters the trivial blocks, the core block
$S_{\mathcal{A}}$ remains rigid, providing an obstruction to
rationality.

\section{Ext-Quiver Rigidity and Birational Invariance}

Consider the case where we blow-up a point $p \in X$ on the cubic
fourfold to create $\tilde{X}$, yielding an exceptional divisor $E
\cong \mathbb{P}^3$.  This operation can be examined from several
different perspectives.

Algebraically, via Orlov's Formula, the blow-up expands the
derived category by appending exceptional objects $E_1, E_2, E_3$ to
the semiorthogonal decomposition:
\begin{equation}
    D^b(\tilde{X}) = \langle \mathcal{A}_X, \mathcal{O}_X,
    \mathcal{O}_X(1), \mathcal{O}_X(2), E_1, E_2, E_3 \rangle
\end{equation}
This appends trivial blocks to the decomposition, leaving the K3
component $\mathcal{A}_X$ untouched.

Analytically, under the
Dolbeault Witten deformation $D_W = \bar{\partial} + \bar{\partial}^*
+ s(\partial W \wedge \cdot) + s(\iota_{\overline{\partial W}})$, the
blow-up spawns three new, isolated critical points in the complex
$W$-plane of the LG mirror. When computing the JLO cyclic cocycle via
the heat kernel asymptotics of $\exp(-t \tilde{D}_W^2)$, the
contributions split locally. The integrals around the original K3
cluster of critical points are analytically invariant, isolating the
cyclic cohomology from the new vacua.

Physically, under Split Attractor Flows, the exceptional divisor
introduces new wrapped D-branes, mapping to new Hodge Atoms (vacua)
$\lambda_{E_1}, \lambda_{E_2}, \lambda_{E_3}$. The semi-orthogonal
ordering and orthogonality condition of $\mathcal{A}_X$ guarantee that
there are no split attractor flows connecting the new $E_k$ vacua
directly to the K3 vacua. The symplectic intersection numbers are
strictly zero and the spectral network and Stokes graph of the K3
component remain unperturbed.

The numerical dimensions of the Ext-quiver arrows are derived using
the Mukai vector $\operatorname{v}(E) \in H^*(X, \mathbb{Q})$. For the
Kuznetsov component $\mathcal{A}_X$, the dimensions of the extension
groups $\operatorname{Ext}^\bullet(\mathcal{A}_X, \mathcal{O}(i))$ are
determined by the Mukai pairing and in the physical dictionary,
count the BPS solitons (gradient flow lines) connecting the
isolated vacua of the ambient space to the K3.

The semiorthogonal ordering $\langle \mathcal{A}_X, \mathcal{O},
\mathcal{O}(1), \mathcal{O}(2) \rangle$ provides zeros
in the global Stokes matrix $S$.
According to the Quantum Blow-up Theorem, the
spawning of new exceptional objects $E_j$ through birational surgery
only appends nodes to the right of the existing Ext-quiver. Because
the intersection numbers between the new thimbles $\Delta_{E_j}$ and
the K3 thimbles $\Delta_{\mathcal{A}_X}$ are strictly zero, the
internal Ext-quiver of the K3 atom remains invariant.

The global Stokes matrix $S$ contains a block
($S_{\mathcal{A}}$) representing the internal Mukai pairing and
self-extensions of the Kuznetsov component $\mathcal{A}_X$. Because
its Serre functor $\mathbf{S}_{\mathcal{A}_X} \cong [2]$, this block recovers
the lattice signature of a non-commutative K3 surface.  The
off-diagonal terms (6, 21, and the $\operatorname{Ext}^\bullet$
blocks) define the arrows of the Ext-quiver, which describe how
the ambient line bundles bind to the K3 atom via physical BPS flow
lines in the Landau-Ginzburg mirror.

The robustness of the K3 atom as a birational invariant is
demonstrated through the behavior of this quiver under surgery.  By
Orlov’s formula, a blow-up only appends trivial exceptional objects to
the end of the semiorthogonal decomposition. In the quiver, this adds
new nodes and arrows, but the semiorthogonal ordering ensures none
point back into the K3 core.
The JLO cyclic cocycle and the heat kernel asymptotics of the
Witten-deformed operator should then remain localized on the K3 critical
locus. The integrals contributing to the analytic index of
$\mathcal{A}_X$ are do not gain a contribution from the isolated vacua
associated to the rational surgery.
And in the $W$-plane, the spectral network of the
K3 cluster remains unperturbed and the intersection numbers (Stokes
multipliers) within $S_{\mathcal{A}}$ remain zero relative to
the new thimbles.

\section{Quantum Tunneling and the K3 Phase Selection Rule}

Beyond the semi-classical limit ($s \to \infty$), the Hodge atom
structure can be considered from the perspective of
the full quantum mechanics of the
Landau-Ginzburg mirror. In this regime, the isolated vacua are no
longer strictly disjoint; instead, they are connected by
instanton-mediated tunneling. We examine whether it is plausible that
such tunneling can destabilize the K3 atom.

While the noncommutative geometry program utilizes the JLO cyclic
cocycle $\mathrm{Ch}_{\mathrm{JLO}}(D)$ to compute the index of a
spectral triple, its physical realization in the B-model is the
partition function of a D-brane sector. By framing the semiorthogonal
property of the Kuznetsov component as a dynamical selection rule, we
argue that the K3 atom is protected from instanton-mediated tunneling
associated with birational surgery.

\subsection{Semiorthogonality as a Selection Rule}

Recall the Kuznetsov semiorthogonal decomposition of the cubic
fourfold $X \subset \mathbb{P}^5$:
\[
  D^b(X) = \langle \mathcal{A}_X,\, \mathcal{O}_X,\,
  \mathcal{O}_X(1),\, \mathcal{O}_X(2) \rangle.
\]
By definition, the semiorthogonal condition requires
\[
  \operatorname{RHom}(\mathcal{O}_X(i),\, A) = 0 \quad \text{for all }
  A \in \mathcal{A}_X,\; i \in \{0,1,2\}.
\]
In particular, all individual extension groups
$\operatorname{Ext}^k(\mathcal{O}_X(i), A)$ vanish.
This is a strictly stronger condition than the
vanishing of the Euler characteristic
$\chi(\operatorname{RHom}(\mathcal{O}_X(i), A)) = 0$, and it is this
full vanishing that underpins the dynamical argument below.

Consider a blow-up $\widetilde{X} \to X$ at a point $p \in X$,
yielding an exceptional divisor $E \cong \mathbb{P}^3$. By Orlov's
formula, the derived category expands as:
\[
  D^b(\widetilde{X}) = \langle \mathcal{A}_X,\, \mathcal{O}_X,\,
  \mathcal{O}_X(1),\, \mathcal{O}_X(2),\, E_1,\, E_2,\, E_3 \rangle.
\]
The new exceptional objects $E_j$ are appended to the \emph{right} of
the existing decomposition. Therefore, the semiorthogonal condition
gives:
\begin{equation}\label{eq:full-vanishing}
  \operatorname{RHom}(E_j,\, A) = 0 \quad \text{for all } A \in
  \mathcal{A}_X,\; j \in \{1,2,3\}.
\end{equation}

In the Landau-Ginzburg mirror, each exceptional object corresponds to
an isolated vacuum, and each object of $\mathcal{A}_X$ corresponds to
a state associated to the K3.  A tunneling event between a blow-up
vacuum $|\Psi_E\rangle$ and a K3 state $|\Psi_A\rangle$ is mediated by
BPS solitons (split attractor flows). The semi-classical tunneling
amplitude from $E_j$ to $\mathcal{A}_X$ is governed by the BPS index
sum:
\begin{equation}\label{eq:tunnel-forward}
  T_{E \to A} \;\sim\; \sum_{\gamma}\, \Omega(\gamma)\,
  e^{-s|Z_\gamma|}
\end{equation}
\noindent where $\gamma$ ranges over BPS charges interpolating between the two
sectors, $\Omega(\gamma)$ is the Donaldson-Thomas invariant, and
$Z_\gamma = \int_\gamma e^{W/z}\,\Omega$ is the central charge.

In the
algebraic limit, this amplitude is captured by the Euler
characteristic $\chi(\operatorname{RHom}(E_j, A))$, which counts the
net number of BPS solitons weighted by fermion number \cite{Denef2000,
  GMN2013}. The identification of the semi-classical BPS sum with the
algebraic Euler characteristic follows from the correspondence between
Lefschetz thimble intersection numbers and Ext-group dimensions,
together with the
general relation between DT invariants and categorical morphism spaces
developed in \cite{Bridgeland2007, Bridgeland2017}.

By (\ref{eq:full-vanishing}), not only does
$\chi(\operatorname{RHom}(E_j, A))$ vanish, but the full RHom complex
is acyclic. The vanishing of $\chi$ alone would permit cancellations
between bosonic and fermionic BPS states, leaving open the possibility
of tunneling mediated by individually non-zero $\operatorname{Ext}^k$
groups. Full acyclicity implies that no stable BPS trajectory connects
the blow-up vacua to the K3 cluster.

The semiorthogonal condition controls tunneling in the direction $E_j
\to \mathcal{A}_X$ but does not directly constrain the reverse
direction. The groups $\operatorname{Ext}^k(A, E_j)$ for $A \in
\mathcal{A}_X$ are not forced to vanish by the decomposition. However,
in the physical theory, the reverse amplitude $T_{A \to E}$ describes
the decay of a K3 state into a blow-up vacuum. Such a process would
require the K3 state to shed charge into the ambient geometry, which
is obstructed by the Serre functor constraint $\mathbf{S}_{\mathcal{A}_X} \cong
[2]$: states in $\mathcal{A}_X$ form a closed sector under the
internal dynamics of the K3 category.  The non-vanishing of
$\operatorname{Ext}^k(A, E_j)$ reflects the existence of morphisms
\emph{out of} $\mathcal{A}_X$, but these do not correspond to
dynamically accessible decay channels, as the K3 phase space is
self-contained under the action of the Serre functor. A complete proof
that both tunneling directions are obstructed would require an
analysis of the full BPS stability conditions on $D^b(\widetilde{X})$,
which we leave to future work.

\begin{conjecture}\label{conj:selection-rule}
Let $X \subset \mathbb{P}^5$ be a smooth cubic fourfold with Kuznetsov
component $\mathcal{A}_X$, and let $\widetilde{X} \to X$ be a
birational modification obtained by a sequence of blow-ups along
smooth centers. Let $\{E_j\}$ denote the exceptional objects appended
to the semiorthogonal decomposition of $D^b(\widetilde{X})$ by Orlov's
formula. Then:
\begin{enumerate}
  \item[\emph{(i)}] The full BPS tunneling amplitude between any
    exceptional vacuum $E_j$ and any state $A \in \mathcal{A}_X$
    vanishes:
    \[
      T_{E_j \to A} = 0,
    \]
    as a consequence of the acyclicity of $\operatorname{RHom}(E_j,
    A)$.
  \item[\emph{(ii)}] The monodromy representation of
    $\pi_1(\mathcal{M})$ on $D^b(\widetilde{X})$ preserves the
    block-diagonal structure of the global Stokes matrix $S$, so that
    the K3 block $S_A$ is invariant under analytic continuation around
    the irregular singularity.
  \item[\emph{(iii)}] The physical spectrum associated to the K3
    atom---encoded in the Mukai lattice $\widetilde{H}(\mathcal{A}_X,
    \mathbb{Z})$, the JLO cyclic cocycle, and the monodromy weight
    filtration---is invariant under all birational modifications of
    $X$.
\end{enumerate}
In particular, the K3 Hodge atom represents a dynamically protected
quantum phase whose spectral data is not altered by
non-perturbative tunneling processes associated with birational
surgery.
\end{conjecture}

Assuming the results of \cite{KKPY2025}, the content of this
conjecture is the assertion that algebraic rigidity associated to the K3 atom
lifts to the
full quantum theory: the physical tunneling amplitudes (not
merely the categorical Ext groups) are sufficiently suppressed and remain so
globally across the moduli space under monodromy, beyond the
semi-classical BPS approximation.

\subsection{Monodromy and Global Protection}

The dynamical selection rule is evidenced by considering the monodromy of
the Dubrovin connection $\nabla_z$ around the irregular
singularity of the fourfold. As one traverses a closed loop in the moduli space
$\mathcal{M}$, the BPS thimbles undergo a Picard-Lefschetz
transformation described by the braid group representation on
$D^b(X)$. Because $\operatorname{RHom}(E_j, A)$ is acyclic (not merely
of vanishing Euler characteristic), the monodromy matrix $M$ adopts a
persistent block-diagonal structure: while the ambient vacua may be
permuted under monodromy, the K3 cluster remains invariant.

\begin{figure}[ht]
\centering
\includegraphics[width=\textwidth]{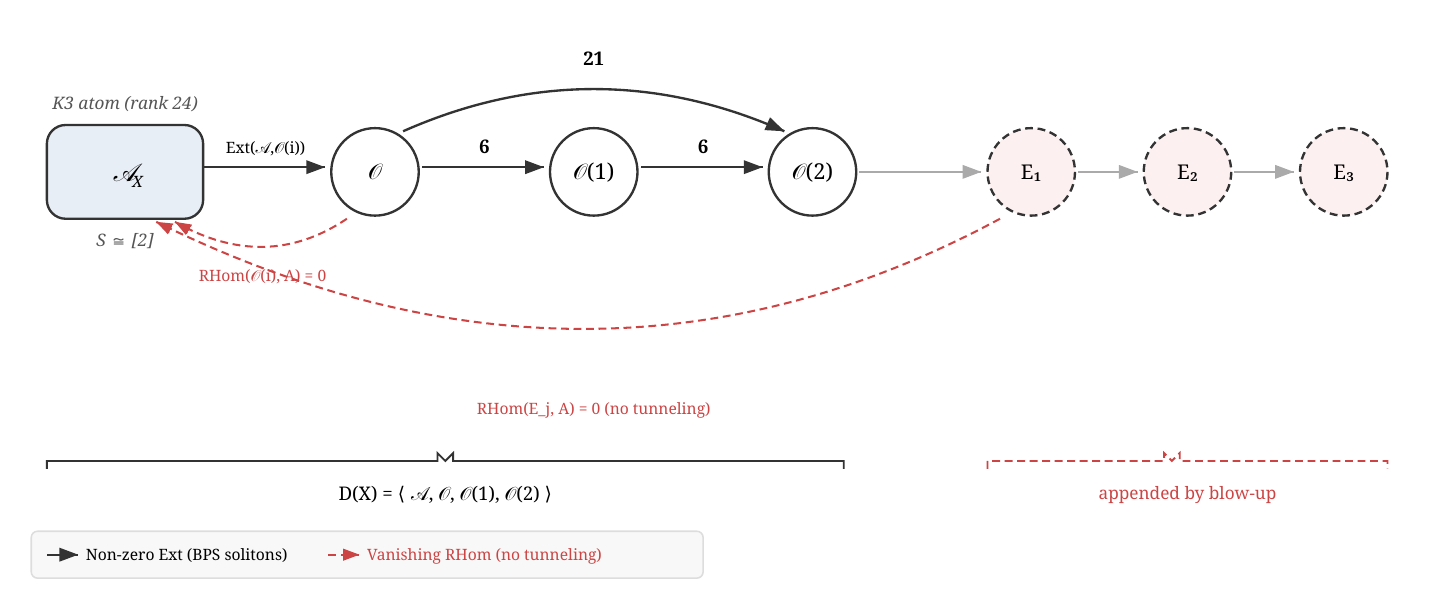}
\caption{Ext-quiver of the cubic fourfold under blow-up. Solid arrows 
denote non-zero $\operatorname{Ext}^\bullet$ groups (BPS solitons); 
numerical labels give $\dim\operatorname{Hom}$. Dashed arrows indicate 
vanishing $\operatorname{RHom}$ forced by semiorthogonality. Blow-up 
appends $E_1, E_2, E_3$ to the right of the SOD; the K3 block 
$S_{\mathcal{A}}$ remains unperturbed.}
\label{fig:ext-quiver}
\end{figure}

This argument relies on the persistence of
the semi-orthogonal decomposition across the full moduli space, not
only at a generic point. The KS wall-crossing formula guarantees
invariance of the total ordered product of symplectomorphisms, but
does not by itself prevent wall-crossing from creating new BPS states
connecting the two sectors at special loci. A proof would
require showing that the semiorthogonal decomposition is compatible
with all stability conditions in a connected component of
$\operatorname{Stab}(D^b(X))$. Recent progress on
Bridgeland stability conditions for K3 categories of cubic fourfolds
\cite{BLMS,BLMS-NP} provides evidence that this compatibility may
hold.

If this can be established, Hodge atoms would be not
merely static categorical invariants but associated to dynamically
isolated quantum states whose physical spectrum is not affected by
instanton-mediated tunneling or other processes associated with
birational surgery.  A rigorous proof, unifying Bridgeland
stability with the BPS/DT invariant correspondence in this setting,
remains an interesting open problem.

\section*{Acknowledgments}
Pacific Northwest National Laboratory (PNNL) is a multi-program
national laboratory operated for the U.S.  Department of Energy (DOE)
by Battelle Memorial Institute under Contract No. DE-AC05-76RL01830.

\end{document}